# Survival Analysis of Young Triple-Negative Breast Cancer Patients

M. Mehdi Owrang O, [1*]Fariba Jafari Horestani,[1] Ginger Schwarz [1]

[1]Department of Computer Science, American University, Washington, DC 20016

*Corresponding author: owrang@american.edu

# Abstract

Breast cancer prognostication is a vital element for providing effective treatment for breast cancer patients. Breast cancer incidence increases with age, with most women diagnosed after the age of 40. However, breast cancer is rare in young women where fewer than five percent of all breast cancers diagnosed in the U.S. occur in women under 40. Previous and current studies suggest that there is poor prognosis in younger women than in older women. In addition, the incidence of breast cancer and prognosis in younger women differ according to ethnicity.

Different types of breast cancer can be identified based on the existence or lack of certain receptors (i.e., estrogen, progesterone, HER2 receptors). Triple-negative breast cancer (TNBC) is characterized by a lack of estrogen receptors (ER), progesterone receptors (PR) and human epidermal growth factor receptor 2 (HER2) expressions. Existing studies suggest that TNBC patients tend to have worse prognosis compared to non-TNBC counterparts. TNBC represents roughly 15% of all breast cancer cases. A higher incidence of TNBC has been seen in younger patients. TNBC has been demonstrated to carry poor prognosis, but whether there exists any age-related variation in TNBC outcomes has yet to be elucidated.

Given the poorer prognosis of TNBC, especially in younger patients, cancer-related outcomes must be estimated accurately. Several factors responsible for the poor clinical outcomes observed in TNBC, including age, race/ethnicity, tumor grade, tumor size, lymph node status, among others, have been studied extensively. Available research data are not conclusive enough to make a convincing argument for or against a biological or clinical difference in TNBC patients based on these factors. This study was designed to investigate the effects of younger age on breast cancer survivability among TNBC patients utilizing population-based Surveillance, Epidemiology, and End Results (SEER) data to confirm whether a younger age factor has prognostic significance.

# 1. Introduction

Breast cancer is the most common female cancer in the US, the second most common cause of cancer death in women [American Cancer Society, 2021], and the main cause of death in women ages 40 to 59 [Siegel 2012].

In 2021, an estimated 281,550 new cases of invasive breast cancer are expected to be diagnosed in women in the U.S., along with 48,100 new cases of non-invasive (in situ) breast cancer [Breastcancer.org]. About 43,600 women in the U.S. were expected to die in 2021 from breast cancer. The lifetime probability of developing breast cancer is one in six overall (one in eight for invasive disease) [American Cancer Society 2021, Siegel 2012, Mohammad Mehdi Owrang O et al., 2023].

Breast cancer incidence is increasing while mortality is declining in many high-income countries [Torre 2016]. The last decade has witnessed a revolution in the understanding of breast cancer, with new classifications proposed that have significant prognostic value and provide guidance on treatment options [Reis-Filho 2008, Boyle 2012].

Medical prognostication is an evaluative component of medicine that encompasses the science of estimating the complication and recurrence of disease and predictive survival of patients [Ohno 2001]. Medical prognosis plays an increasing role in health care outcomes. Many factors, including tumor grade, tumor size, and lymph node status may influence or correlate with prognosis for breast cancer patients.

Different types of breast cancer can be identified based on the existence or lack of certain receptors (i.e., estrogen, progesterone, HER2 receptors) [Onitilo 2009]. Triple-negative breast cancer (TNBC) is characterized by a lack of estrogen receptors (ER), progesterone receptors (PR) and human epidermal growth factor receptor 2 (HER2) expressions [Anders 2008, Dawson 2009, Tim 2011, Boyle 2012, Chen 2012, Lehmann 2014, Wilson 2016,]. Previous and current studies suggest that TNBC patients tend to have worse prognosis compared to non-TNBC counterparts [Basu 2008, Agarwal 2012, Rahmani 2012]. In addition, those studies suspect that the incidence of breast cancer, and prognosis in TNBC women, differ according to age. Age at diagnosis is an important prognostic factor for breast cancer in general; however, its specific significance in triple negative breast cancer (TNBC) is unclear. There is little existing data on the prognostic value of age in TNBC-[Dent 2007, Clarke 2012, Tariq 2014].

Existing works on TNBC are mainly focused on the comparisons of the TNBC and non-TNBC with respect to the clinical, pathological, histological, prognostic features, and outcome associated with these two breast cancer subtypes. Available research data is not conclusive enough to make a convincing argument for or against a biological or clinical difference in TNBC patients based on these factors.

The aim of our study is to see if younger age is a main determinant of long-term survival in women with TNBC tumor type. We present an analysis of the prognosis for the younger TNBC patients, based on the age, age within ethnicity/race, two years survival, five years survival, NPI prognosis (Nottingham Prognostic Index) [NPI – Albergaria 2011], and age within marital status. All experiments are done using SPSS Statistical Package-Survival Analysis [Hui Bian] with the breast cancer data set from the National Cancer Institute's SEER [SEER].

## 2. Breast Cancer Prognosis

Medical prognosis is a field in medicine that encompasses the science of estimating the complication and recurrence of disease and to predict survival of patients [Junsdottir 2008, Gupta 2011, Kharva 2012]. Survival analysis is a field in medical prognosis that deals with the application of various methods to estimate the survival of a patient suffering from a disease.

### 2.1 Breast Cancer Prognostic Factors

A prognostic factor may be defined as a measurable variable that correlates with the natural history of the disease. The prognostic factors used in the prediction of survival of breast cancer can be separated into two categories: chronological (based on the amount of time present, i.e., Stage of Cancer), or biological (based on the potential behavior of the tumor, i.e., Histological Grade) [Bradley 2007, Soerjomataram, 2008, Maskarinec 2011, Costanza 2012].

Lymph node status, tumor size, and histological grade are among the prognostic factors in use today [Delen 2005, Bradley 2007, Soerjomataram 2008, Maskarinec 2011, Costanza 2012]. Lymph node status is a time-dependent factor and is directly related to survival. One of the most significant prognostic factors in breast cancer is the presence or absence of axillary lymph node involvement, which is usually assessed at the time of surgery using sentinel lymph node biopsy or axillary dissection [Bradley 2007]. Macrometastases (>0.2 cm in size) have clearly been shown to have prognostic significance.

Survival is inversely related to the size of the tumor, for both TNBC and non-TNBC. The probability of long-term survival is better with smaller tumors than with larger tumors [Soerjomataram 2008, Maskarinec 2011, Costanza 2012]. Tumor size has long been recognized as an independent prognostic factor and as a predictor of axillary node status, with larger tumors being associated with a worse prognosis and an increased likelihood of nodal metastasis. Bellacchia et al. [Bellaachia 2006] have used the Weka mining tool and ranked the survivability attributes. The result indicates that "Extension of tumor" has a higher rank than the "Tumor size".

Histological grade is highly correlated with long term survival. Patients with a grade I tumor have a much better chance of surviving than patients with grade III tumors [Breast Cancer 2012]. Delen et al. [Delen 2005] also conducted sensitivity analysis on artificial neural networks model gain insight into the relative contribution of the independent variables to predict survivability. The sensitivity results indicated that the prognosis factor "Grade" is by far the most important predictor, which is consistent with the previous research, followed by "Stage of Cancer", "Radiation", and "Number of Primaries" [Delen 2005]. Why these prognostic factors are more important predictors than the other is a question that can only be answered by medical professionals and further clinical studies.

Other factors include the estrogen and progesterone receptor levels, as well as human epidermal growth factor 2 expression in the tumor tissue, and genetic properties of the cancer cells (these properties, which are sometimes called biomarkers, can be determined by specific lab and imaging tests).

## 2.2 Breast Cancer Prognostic Tool: Nottingham Prognostic Index (NPI)

The Nottingham Prognostic Index (NPI) [NPI, Albergaria 2011] is a prognostic tool used, widely in Europe, to determine prognosis following surgery for breast cancer. Its value is calculated using three pathological criteria: the size of the tumor; the number of involved lymph nodes; and the grade of the tumor.

The index is calculated using the formula: NPI = [0.2 x S] + N + G, Where:
**S** is the size of the tumor in centimeters,
**N** is the number of lymph nodes involved: 0 =1, 1-3 = 2, >3 = 3,
**G** is the grade of tumor: Grade I =1, Grade II =2, Grade III =3

The interpretation is as follows

| Score | 5-Year Survival | Prognosis |
|---|---|---|
| $\geq 2.0$ to $\leq 2.4$ | 93% | Excellent Prognosis |
| $> 2.4$ to $\leq 3.4$ | 85% | Good Prognosis |
| $> 3.4$ to $\leq 5.4$ | 70% | Moderate Prognosis |
| $> 5.4$ | 50% | Poor Prognosis |

With NPI, it is possible to group patients with respect to prognosis as those with "Excellent prognosis", "Good prognosis", etc. as defined above. That is why we have decided to do an exploratory analysis of breast cancer data sets using NPI and to group the patients based on their prognosis and age categories to better understand the correlation between prognosis and age.

# 3. Triple Negative Breast Cancer (TNBC) Related Works

## 3.1 TNBC Characteristics

TNBC accounts for 10%–20% of all breast cancer cases. Although breast cancer is one of the most frequent malignancies, histopathological features, and long-term outcome of breast cancer, especially TNBC type, are already uncertain. Triple negative tumors typically are the most common breast cancer subtype in BRCA1 carriers. Compared to other breast carcinoma subtypes, TNBC correlates with more aggressive characteristics such as larger tumor size, higher histological grade, more positive lymph nodes, advanced stage, and younger age. Additionally, it is consequently found more often in pre-menopausal women and is diagnosed more frequently in African American/Hispanic women [Rhee et al. 2008, Stead 2009, Lara-Medina 2011, Clarke 2012, Brewster 2014]. Taken together, these adverse factors may be a major reason that TNBC patients suffer from poorer overall survival (OS), breast-cancer-specific survival (BCSS) and relapse-free survival (RFS) reported for this disease. [Nofech-Mozes et al. 2009, Agarwal 2016].

Review of different studies showed the main risk factors of TNBCs include age (age at diagnosis < 50 years, young age at menarche, young age at time of first birth), race/ethnicity (African American ethnicity, Hispanic, White) among other factors including high body mass index, high parity, lack of breast feeding, and some biological features such as high Ki67, and presence of P53 mutations [Dolle 2009, Huo 2009, Stead 2009]. In a univariate analysis studied by Lara-Medina [Lara-Medina 2011], TNBC was associated with younger age (49.2 years vs 52.2 years; $P < .001$), premenopausal status (Odds Ratio (OR), 0.72; 95% CI, 0.58-0.88; $P = .002$), and advanced disease stage (stages III and IV vs stages I and II: OR, 1.60; 95% CI, 1.33-2.04; $P < .001$). The median age at diagnosis of Hispanic women with breast cancer was 11 years younger than the average age reported in the United States. The prevalence of TNBC in this study population was higher than that reported in white women with breast cancer.

In the population-based Carolina Breast Cancer Study (CBCS) [Carey et. al 2006], in a cohort study of 1601 breast cancer patients (180 triple-negative), the mean age at diagnosis was younger for the triple negative group (53 years versus 58 years). The triple-negative breast cancer patients were also more likely to be grade III and have a greater mean tumor size. Bauer and Colleagues [Baur 2007] reported that compared with women with non-TNBC, the odds of a woman with TNBC being under the age of 40 years was 1.53. In another study, Lee and Colleagues [Lee 2010] analyzed breast cancer subtypes using Korean Breast Cancer Society Registration Program data to compare clinical features and prognosis for triple-negative breast cancer (TNBC). A cohort of 26,767 breast cancer patients were used. TNBC correlated with younger age, characteristics that are more aggressive, and poor overall survival and breast cancer specific survival.

## 3.2 TNBC vs non-TNBC

There have been several retrospective studies comparing the TNBC and non-TNBC patients [Agarwal 2016, Tariq 2014]. The TNBC subtype generally carries a worse prognosis than its non-TNBC counterpart does. In Canadian study, involving a large population by Dent and Colleagues [Dent 2007], women with TNBC had an increased risk of death with hazard ratio of 3.2 compared with women with non-TNBC. Rhee et al. found that the four-year survival of TNBC patients was 85.5%, compared to 94.2% in non-TNBC patients [Rhee et al. 2008]. TN tumors were significantly larger, had frequent p53 overexpression, higher Ki67 values, and a higher nuclear grade in comparison with non-TN tumors. Patients with non-TN tumors had a significantly better Disease-Free Survival (DFS) than those with TN tumors. The OS rate of patients with TN tumors was significantly lower than those with non-TN tumors. [Nishimura 2008]

Differences exist in age at diagnosis for the various breast cancer subtypes. Most triple negative cancers display distinct clinical and pathological characteristics with a high proportion of these tumors occurring at a younger age of onset. Existing work suggested that the patients with TNBC diagnosis were significantly younger than non-TNBC. In a study conducted by Gaurav Agarwal and colleagues [Agarwal 2016], TNBC patients constituted 35.3% (249 of 705) of the entire study cohort. A comparison of TNBC and non-TNBC patients revealed that TNBC patients were younger, and more often premenopausal (47% in TNBC compared with 38.4% in non-TNBC group, $p = 0.03$). Mean tumor size was similar in the two groups ($5.7 \pm 2.9$ cm in TNBC compared with $5.4 \pm 2.8$ in non-TNBC, $p = 0.15$). However, a higher proportion of TNBC patients had lymph node metastases (cN status) at presentation (77.3% in TNBC compared with 69.8% in non-TNBC group, $p = 0.03$). The histological grade III tumor proportion was higher in TNBC (56.4%) compared to non-TNBC group (31.4%, $p = 0.002$).

In a study conducted by [Rahmani 2012], a total of 86 of 546 final included participants with breast cancer were identified as having TNBC (15.8%). The patients with TNBC diagnosis were significantly younger than non-TNBC group and family history of breast cancer was more prevalent in former group. Regarding histopathological features, medullary features were more prevalent in TNBC group, while other features were similarly revealed in both groups. With respect to tumor grading, TNBC group was graded higher than non-TNBC group. Grade III tumors were reported in 51.1% of the TNBC patients, but in 15.9% of another group. In addition, the tumor stage was significantly higher in the TNBC group. Tumor size greater than 50 mm was observed in 18.6% of the TNBC and 14.8% of non-TNBC groups. Lymph nodes involved in TNBC group were more prevalent compared with non-TNBC patients (82% vs 67%) [Rahmani 2012].

### 3.3 Young TNBC Related Works

In the current literature, there are several studies that have looked at the prognostic impact of age in patients with triple-negative breast cancer [Kassam 2009, Lee 2011, Liedtke 2013, Liu 2014, Sittig 2017, Ryu 2017, Tano 2017, Otoukesh 2018, Morante 2018, Dai 2019]. In general, researchers identified the characteristics of patients with TNBC as associated with distinct clinical and pathologic characteristics such as, younger age at diagnosis, higher tumor grade, higher cancer stage, and have poorer prognosis than those diagnosed with non-TNBC.

Numerous studies have demonstrated that breast cancer in young women (BCY) has unfavorable prognostic features and more unfavorable subtypes. However, few studies have evaluated the effect of subtype disparities on breast cancer prognosis by age, especially for BCY.

In a previous study of patients with metastatic triple-negative breast cancer, age at diagnosis less than 50 years was an independent adverse prognostic factor in multivariate analysis [Kassam 2009].

In "Different prognosis of young breast cancer patients in their 20s and 30s depending on subtype: a nationwide study from the Korean Breast Cancer Society", authors analyzed breast cancer mortality stratified by tumor subtype according to age among patients younger than 50 years. Patients in the 20s group showed worse prognosis. In multivariate analysis for overall survival (OS), the hazard ratio (HR) for patients in the 20s group was higher than that for the 30s and 40s groups, and patients with triple-negative breast cancer (TNBC) showed higher HR than patients with HER-2 or luminal subtype (all p=0.0001). When stratified by subtype, luminal subtype showed significantly worse prognosis in the 20s group than the 30s and 40s groups, whereas TNBC subtype showed no significant difference between women in the age groups of 20s versus 30s, and 20s versus 40s (p = 0.445 and p = 0.592).

In "Effect of Age in Young Women With Stage I-III Triple-Negative Breast Cancer: A Report From the National Cancer Database", authors aimed to characterize the role of age on survival in young women (45 years) with stage I-III triple-negative breast cancer (TNBC). On multivariate analysis, in comparison to patients aged 41-45 years, there was no significant difference in survival compared to the first age group (age 20-25 years), second age group (age 26-30 years), third age group (age 31-35 years), or fourth age group (age 36-40 years). Increasing primary tumor size, increasing number of positive lymph nodes, and African American race were associated with decreased survival across all age groups in multivariate analysis. This data demonstrates that among women 45 years of age with TNBC, younger age is not associated with worse overall survival.

Authors in "Prognostic impact of age at diagnosis in triple negative breast cancer: Analysis of 204 patients from single institution registry", evaluated the prognostic value of age at presentation of 204 TNBC patients diagnosed at Loma Linda University Cancer Center. The Kaplan Meier method was used to calculate survival as well as to calculate it with respect to age. Age was categorized into less than 40 and greater than 40. Age was found to be significantly associated with race. Specifically, there were proportionally larger number of Hispanic patients younger than 40 years compared to non-Hispanic patients (P = 0.002). There was no correlation with age and cancer stage, histology, lymph node status, type of surgery or receipt of radiation.

Authors Liu, Xin, et al, compared clinical pathological characteristics and prognosis of very young (<= 35) and older (> 35) triple-negative breast cancer (TNBC) patients to assess their relevance to TNBC in a younger population [Liu 2014]. As calculated through Kaplan–Meier, their study confirmed that TNBC patients aged <=35 years have poorer disease-free survival (DFS) than those elderly group (5-year DFS). Very Young TNBC patients have showed a high clinical stages and more positive lymph nodes. The overall survival of both young and older TNBC did not have significant differences. In another study [Morante 2018], authors describe the clinicopathological and epidemiological features of triple-negative breast cancer (TNBC) in patients aged ≤35 years. In total, 243 of 2007 cases (12.11%) were very young TNBC patients. The Cox proportional hazard model was used to identify prognostics factors for DFS and OS. A high histological grade was frequent (84.7%). TNBC in very young Peruvian women was characterized by advanced stage at diagnosis. In these patients, nodal involvement was the most important prognostic factor for DFS. It presents distinctive characteristics and poorer outcomes in terms of DFS and OS.

Tano's retrospective review [Tano 2017] identified 194 patients with TNBC. The patient's race, age at diagnosis, stage at diagnosis, treatment, relapse free survival (RFS), overall survival, and survival following confirmed metastatic disease were analyzed. There was no statistical difference in frequency of stage between groups. Although not statistically significant, there was a trend of to worse RFS in African American (AA) patients and in young patients of different races. Younger age at diagnosis did have an inferior median survival as compared to older patients regardless of race or stage (p=0.0476). There was no difference in median survival following confirmation of metastatic disease between races (p=0.3574) or between age groups (p=0.8548); (median survival = 31 months, 2.56-204 months).

The purpose of Liedtke's study [Liedtke 2013] was to assess the prognostic impact of age in patients with triple-negative breast cancer (TNBC). They demonstrated in a large dataset of 1,732 patients with TNBC that young age at diagnosis is an important unfavorable prognostic factor. Five age cohorts (B30, 31–40, 41–50, 51–60, and >60 years) at diagnosis were correlated with clinical/pathological parameters. Univariate and multivariate analyses were used to examine the effect of age on disease-free (DFS), distant disease-free (DDFS), and overall survival (OS). In patients with TNBC, increasing age at diagnosis was inversely correlated with tumor grade (P=0.0001); likelihood of being non-Caucasian (P = 0.0001). Younger patients with TNBC were more often diagnosed with grade III tumors (i.e., patients aged 31–40 and >60 years had 93 and 83%, respectively, P=0.0001).

The sparse research data available on breast cancer in general has shown variable results, with some making a strong case for age as a distinct prognostic factor in younger patients and others failing to support this relationship [Clarke 2012]. It is not well understood whether the prognosis differs between patients who develop TNBC at a younger age versus those who develop it at an older age. Available research data are not conclusive enough to make a convincing argument for or against a biological or clinical difference in TNBC patients based on age at diagnosis [dent 2007, Tariq 2014].

## 4. Young Triple Negative Breast Cancer Survival Analysis

Despite the widespread acknowledgement of the poor clinical outcome of TNBC, the prognostic value of specific biological features of these tumors continues to raise substantial degree of uncertainty and controversy.

Traditionally, several clinicopathological characteristics, such as patient age, tumor size, histological grade, hormone receptor status, her2 status, lymphovascular invasion, and lymph node involvement have been used to determine the prognosis of breast cancer patients. However, there are reports that suggest, except for lymph node involvement, those traditional markers are of little value in predicting the prognosis of TNBC patients [Shen 2012]. The relation between TNBC and lymph node status is less clear. Some reports found no relation, others found a negative relation, and some found a positive relation. The prognostic value of classical pathological variables such as tumor size, tumor grade, lymph node status could be impaired in TNBC [Broukaer 2012]

The information of age, tumor size, lymph node status, clinical stage, histological grade, and pathological types play important roles in judging the prognosis of TNBC [Qiu 2016]. TNBC has been demonstrated to carry poor prognosis but whether there exists any age-related variation in TNBC outcomes has yet to be elucidated [Zhu 2015].

The aim of this study is to analyze the clinical and pathological characteristics, survival, and prognostic factors of the Triple Negative Breast Cancer patients. We will investigate the significance of the younger age on the overall prognosis and survive rate of the TNBC patients. Survival Analysis experiments are to be performed on SEER breast cancer dataset [SEER] and analyzed using Kaplan-Meier survival curve [Kaplan-Meier] of the IBM SPSS statistical package [Hui Bian].

## 4.1 Odds of Survival for the young TNBC Patients Compared to young non-TNBC Patients

We have done many survival analysis experiments using the IBM SPSS Survival Analysis [Bian], Kaplan-Meier survival curve, and MedCalc (Odds Ratio Calculator) [MedCalc_Odds Ratio Calculator] with SEER breast cancer dataset, to compare the survival rates of the young non-TNBC and young TNBC patients.

For the age group, both the younger ($<30$) and older ($>= 30$) non-TNBC patients have a higher odd of survival compared to the TNBC patients. Based on Table 1, for the age group $< 30$, the odds of survival for the non-TNBC patients is 2.56 times as high as that of the TNBC patients (OR=2.56, $P < 0.0001$, Figure 1). Likewise, for the age group $>= 30$, the odds of survival for the non-TNBC patients is 1.89 times as high as that of the TNBC patients (OR=1.89, $P < 0.0001$).

**Table 1.** KM Case Processing Summary data for Age 30
(N of Events=survived, Censored=died

### Case Processing Summary

| Age30 | TNBC | Total N | N of Events | Censored N | Percent |
|---|---|---|---|---|---|
| <30 | No | 5490 | 3703 | 1787 | 32.6% |
|  | Yes | 265 | 223 | 42 | 15.8% |
|  | Overall | 5755 | 3926 | 1829 | 31.8% |
| >=30 | No | 747027 | 554535 | 192492 | 25.8% |
|  | Yes | 22971 | 19400 | 3571 | 15.5% |
|  | Overall | 769998 | 573935 | 196063 | 25.5% |
| Overall | Overall | 775753 | 577861 | 197892 | 25.5% |

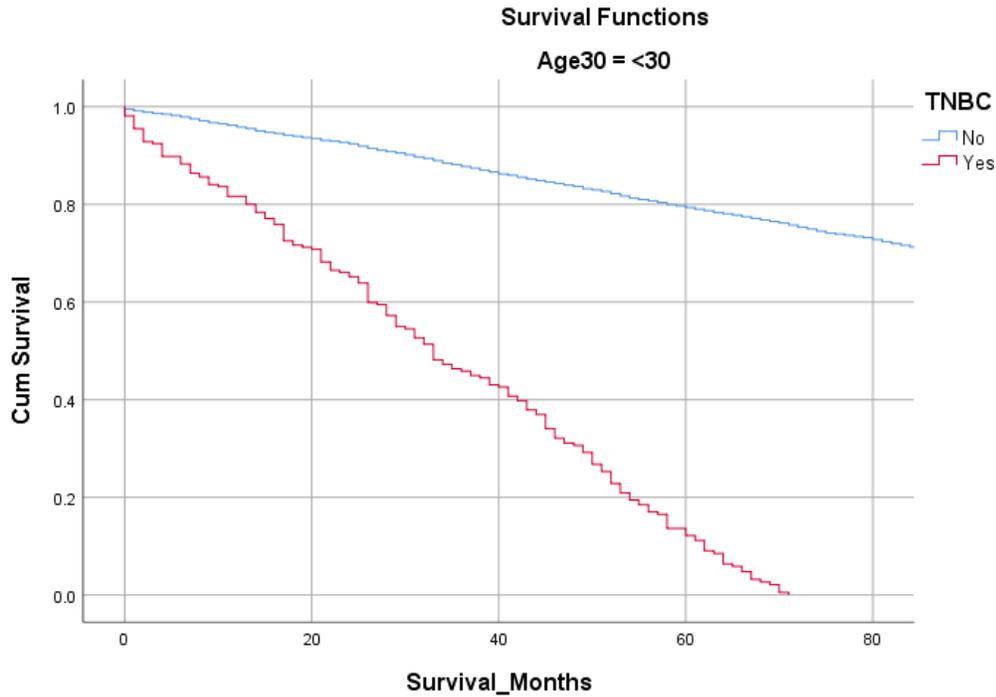

**Figure 1**. Kaplan-Meier survival curve for non-TNBC vs TNBC patients for Age group < 30

## 4.2 Odds of Survival for the Young TNBC Patients

As we noted, existing works on TNBC are mainly focused on the comparisons of the TNBC and non-TNBC, with respect to the clinical, pathological, histological, prognostic features, and outcome associated with these two breast cancer subtypes. Our experiments, represented in section 4.1, strongly concur the findings of the previous and current results that non-TNBC patients (young and old) have better prognosis than the TNBC patients do.

In this section, we compare the survival rate of the TNBC patients based on their age, age and ethnicity/race, age and marital status, two years survival, five years survival, and NPI prognosis. The survival analysis has been done using the IBM SPSS statistical software on SEER breast cancer dataset, Kaplan-Meier Survival Curve, and MedCalc Odds Ratio Calculator.

Our aim in this study is to compare TNBC patients to see whether there exists any age-related variation in TNBC prognosis.

## Younger TNBC Patients' survival

Examining Table 2, we find that the odds of survival for the younger (age of < 30 years) patients are slightly higher than the older (age of >= 30 years) patients (using MedCalc Odds Ratio Calculator, OR=1.324, P = 0.078). However, the difference is not statistically significant (Kaplan-Meier (KM) curve, Figure 2).

**Table 2.** KM Case Processing Summary data for TNBC patients based on Age group.
(N of Events=survived, Censored=died)

**Case Processing Summary**

| Age30 | Total N | N of Events | Censored N | Percent |
|---|---|---|---|---|
| < 30 | 256 | 206 | 50 | 19.5% |
| >= 30 | 22980 | 19417 | 3563 | 15.5% |
| Overall | 23236 | 19623 | 3613 | 15.5% |

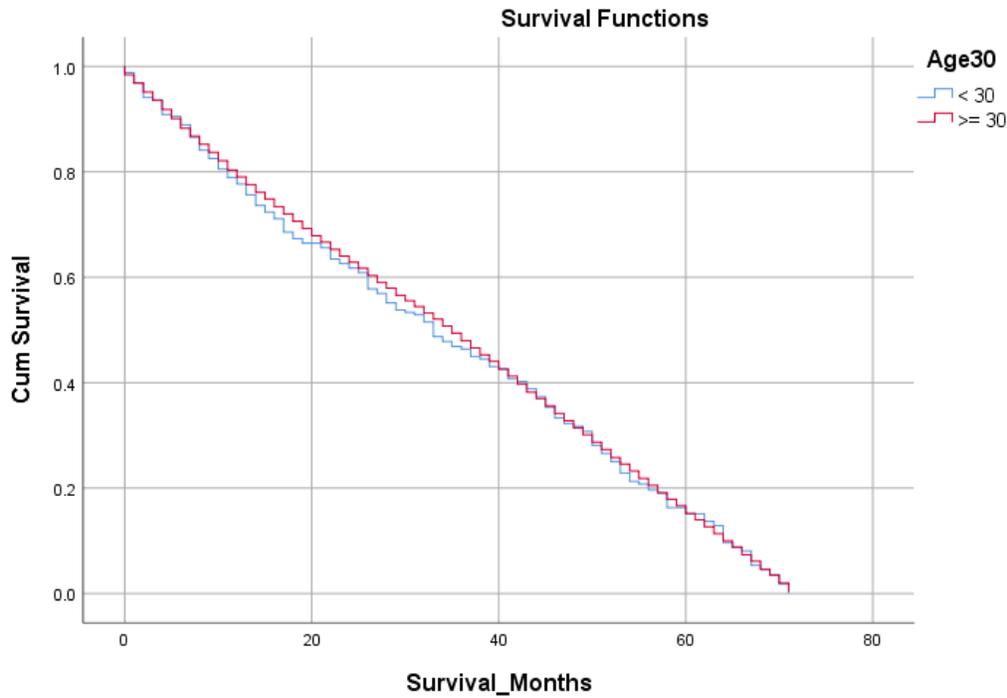

**Figure 2**. KM survival months for TNBC patients by Age

## Younger TNBC patients' survival by Ethnicity/Race

Looking at Table 3, we can find out that the younger TNBC HWhite, NHAPI, and NHWhite (Figure 3) patients have slightly higher survival rate than the older ones whereas the older NHBlack (Figure 4) patients had slightly better survival rate than the younger ones (Figures 3 to 6). However, in all cases, the differences were not statistically significant. In general, the sample sizes were too low to consider the differences significant anyway.

**Table 3.** KM Case Processing Summary data for TNBC patients based on Ethnicity_Race and Age. (N of Events=survived, Censored=died)

**Case Processing Summary**

| Ethnicity_Race | Age30 | Total N | N of Events | Censored N | Percent |
|---|---|---|---|---|---|
| HBlack | >= 30 | 52 | 41 | 11 | 21.2% |
|  | Overall | 52 | 41 | 11 | 21.2% |
| HWhite | < 30 | 64 | 54 | 10 | 15.6% |
|  | >= 30 | 3067 | 2732 | 335 | 10.9% |
|  | Overall | 3131 | 2786 | 345 | 11.0% |
| NHAIAN | < 30 | 4 | 4 | 0 | 0.0% |
| | >= 30 | 178 | 171 | 7 | 3.9% |
| | Overall | 182 | 175 | 7 | 3.8% |
| NHAPI | < 30 | 27 | 17 | 10 | 37.0% |
| | >= 30 | 2095 | 1645 | 450 | 21.5% |
| | Overall | 2122 | 1662 | 460 | 21.7% |
| NHBlack | < 30 | 54 | 47 | 7 | 13.0% |
| | >= 30 | 3830 | 3266 | 564 | 14.7% |
| | Overall | 3884 | 3313 | 571 | 14.7% |
| NHUnk | < 30 | 1 | 1 | 0 | 0.0% |
| | >= 30 | 103 | 80 | 23 | 22.3% |
| | Overall | 104 | 81 | 23 | 22.1% |
| NHWhite | < 30 | 104 | 82 | 22 | 21.2% |
| | >= 30 | 13590 | 11430 | 2160 | 15.9% |
| | Overall | 13694 | 11512 | 2182 | 15.9% |
| Other | < 30 | 2 | 1 | 1 | 50.0% |
| | >= 30 | 65 | 52 | 13 | 20.0% |
| | Overall | 67 | 53 | 14 | 20.9% |
| Overall | Overall | 23236 | 19623 | 3613 | 15.5% |

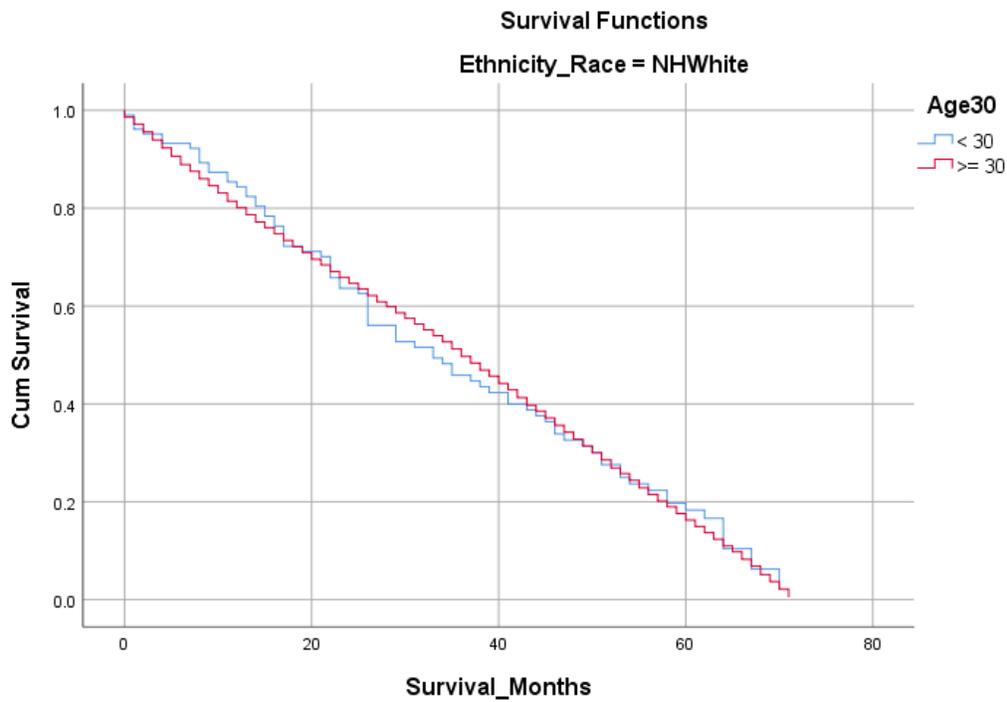

**Figure 3**. KM survival months for TNBC patients by Age for NHWhite Ethnicity_Race.

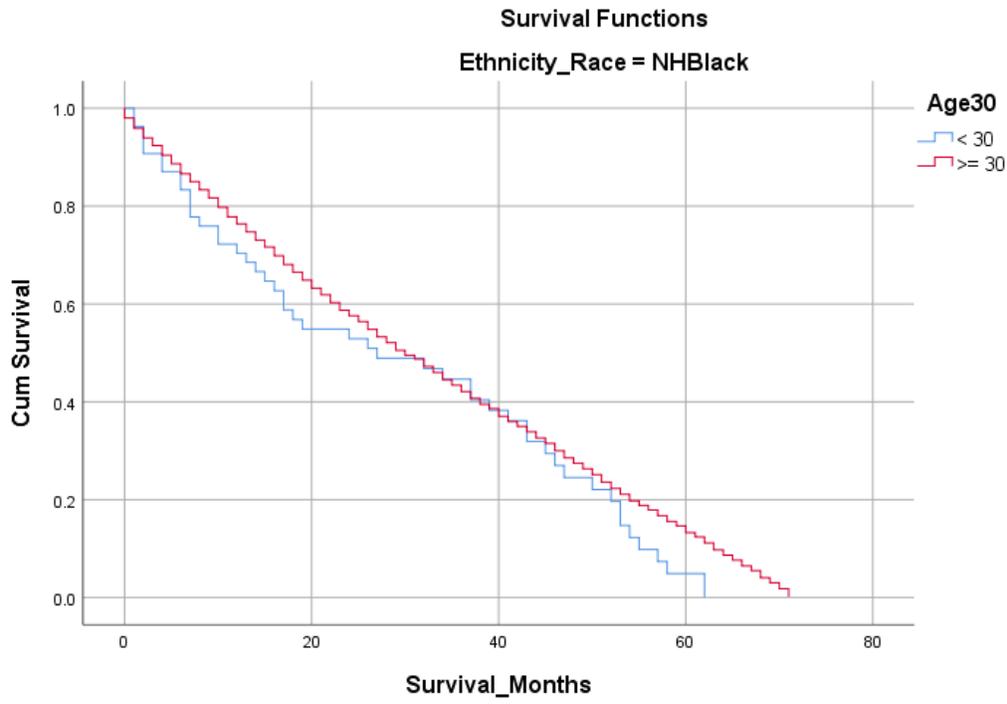

**Figure 4**. KM survival months for TNBC patients by Age for NHBlack Ethnicity_Race.

# Younger TNBC patients' survival by Marital Status

Examining Table 4, we find that the odds of survival for the younger (age of < 30 years) married patients are slightly higher than the older (age of >= 30 years) married patients (OR=1.13, P = 0.67, Figure 5). Likewise, the odds of survival for the younger (age of < 30 years) single patients are slightly higher than the older (age of >= 30 years) single patients (OR=1.48, P = 0.053, Figure 6).

As for the divorced and widowed young patients, the sample sizes were too low to be considered.

**Table 4**. KM Case Processing Summary data for TNBC patients based on Ethnicity_Race and Age. (N of Events=survived, Censored=died)

**Case Processing Summary**

| Marital_Status | Age30 | Total N | N of Events | Censored N | Percent |
|---|---|---|---|---|---|
| Divorced | < 30 | 2 | 2 | 0 | 0.0% |
|  | >= 30 | 2641 | 2225 | 416 | 15.8% |
|  | Overall | 2643 | 2227 | 416 | 15.7% |
| Married | < 30 | 89 | 74 | 15 | 16.9% |
|  | >= 30 | 12137 | 10287 | 1850 | 15.2% |
|  | Overall | 12226 | 10361 | 1865 | 15.3% |
| Other | < 30 | 18 | 16 | 2 | 11.1% |
|  | >= 30 | 1553 | 1354 | 199 | 12.8% |
|  | Overall | 1571 | 1370 | 201 | 12.8% |
| Single | < 30 | 145 | 112 | 33 | 22.8% |
|  | >= 30 | 3778 | 3151 | 627 | 16.6% |
|  | Overall | 3923 | 3263 | 660 | 16.8% |
| Widowed | < 30 | 2 | 2 | 0 | 0.0% |
|  | >= 30 | 2871 | 2400 | 471 | 16.4% |
|  | Overall | 2873 | 2402 | 471 | 16.4% |
| Overall | Overall | 23236 | 19623 | 3613 | 15.5% |

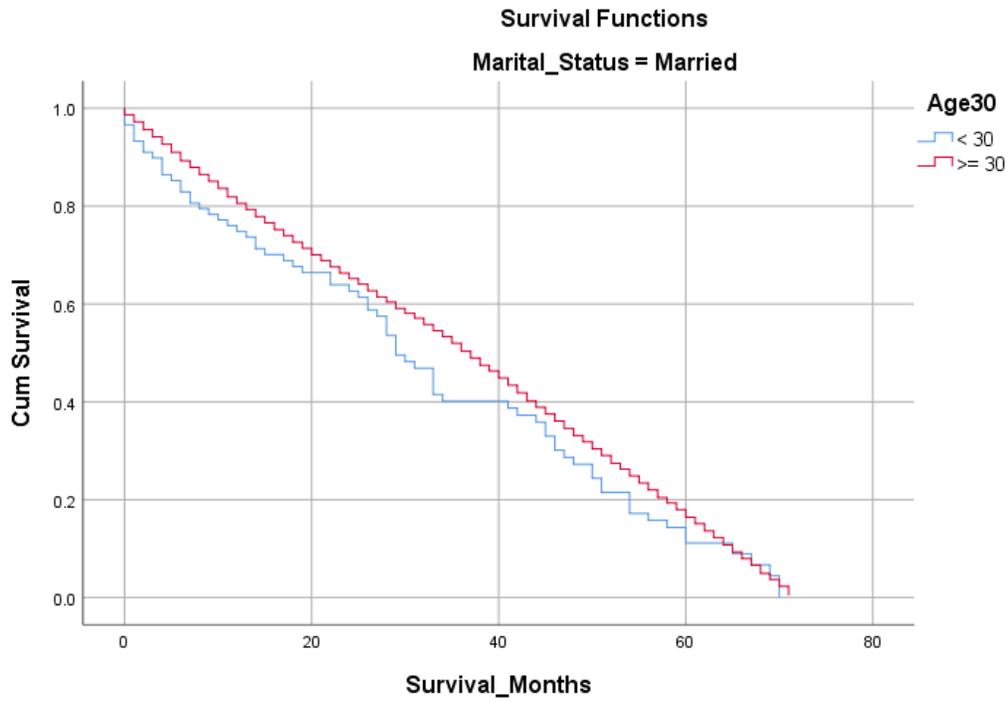

**Figure 5**. KM survival months for TNBC patients by Age and Married patients.

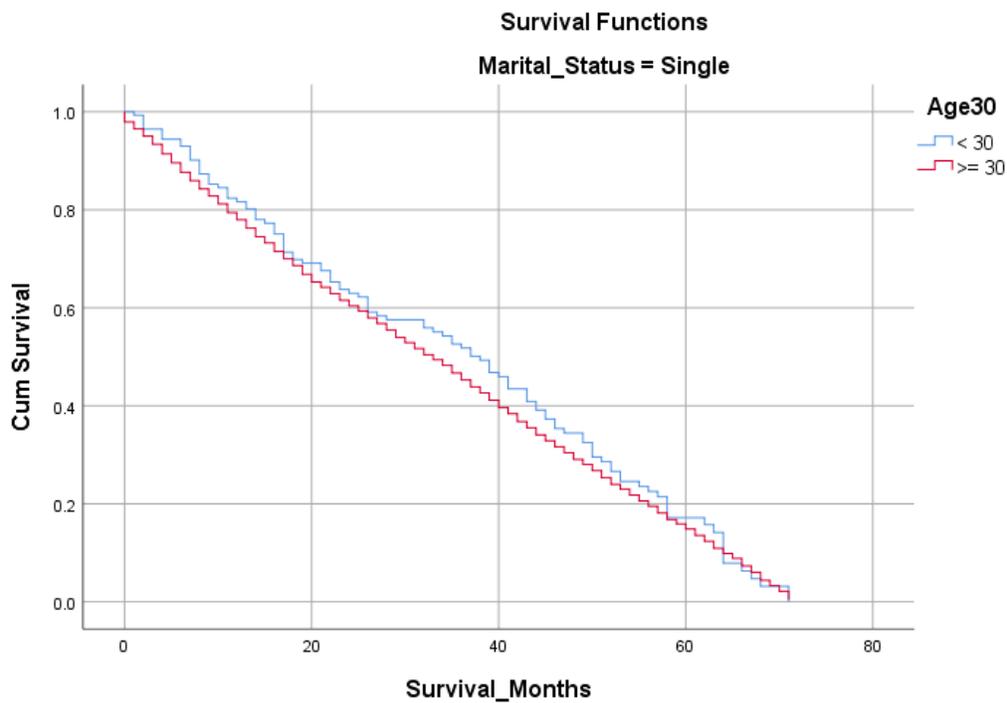

**Figure 6**. KM survival months for TNBC patients by Age and Single patients.

## Younger TNBC patients and Two Years Survival

Considering Table 5, we find that the odds of survival for less than two years for the younger (age of < 30 years) patients are slightly higher than the older (age of >= 30 years) patients (OR=1.2, P = 0.38, Figure 7). Similarly, for the greater than two years survival, the younger patients have slightly higher survival rate than the older patients (OR=1.4, P=0.12, Figure 8). However, in both cases, the differences are not statistically significant.

**Table 5**. KM Case Processing Summary data for TNBC patients on Age and Two Years Survival. (N of Events=survived, Censored=died)

### Case Processing Summary

| TwoYearsSurvival | Age30 | Total N | N of Events | Censored N | Percent |
|---|---|---|---|---|---|
| <2Years | < 30 | 121 | 98 | 23 | 19.0% |
|  | >= 30 | 9507 | 7977 | 1530 | 16.1% |
|  | Overall | 9628 | 8075 | 1553 | 16.1% |
| >=2Years | < 30 | 135 | 108 | 27 | 20.0% |
|  | >= 30 | 13472 | 11439 | 2033 | 15.1% |
|  | Overall | 13607 | 11547 | 2060 | 15.1% |
| Overall | Overall | 23236 | 19623 | 3613 | 15.5% |

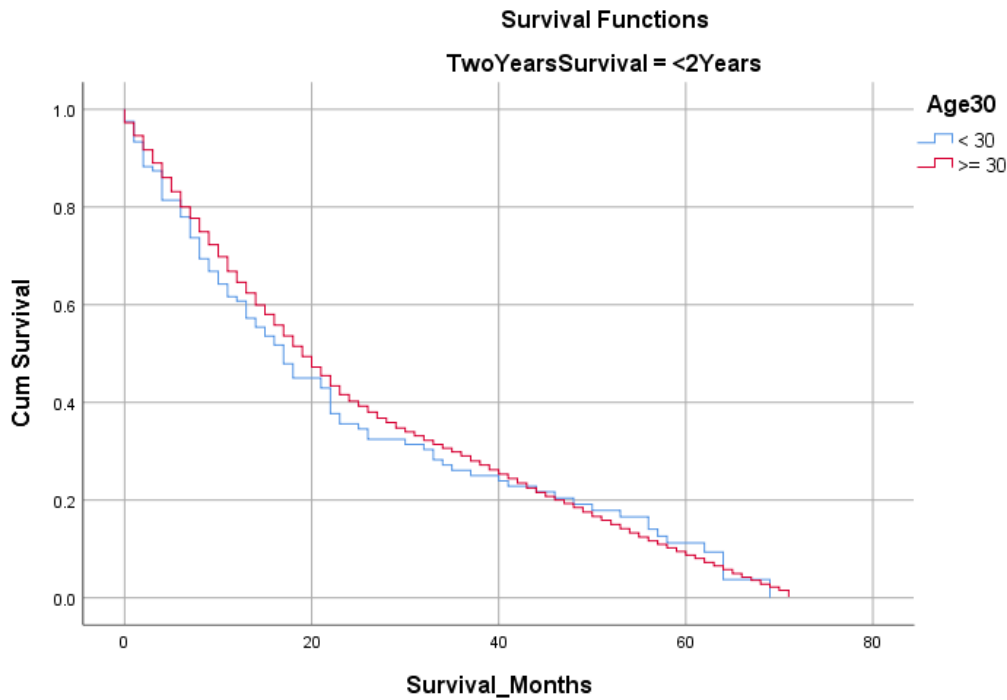

**Figure 7**. KM survival months for TNBC patients by Age and Two_Years_Survival <2 years.

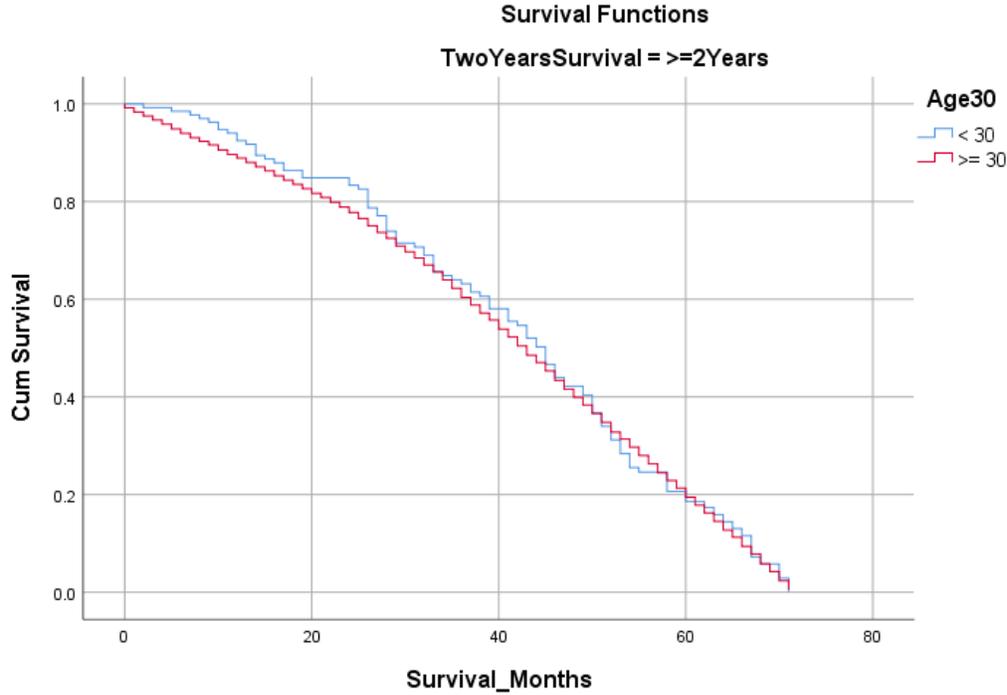

**Figure 8**. KM survival months for TNBC patients by Age and Two_Years_Survival >= 2 years.

## Younger TNBC patients and Five Years Survival

Considering Table 6, we find that the odds of survival for five years for the younger (age of < 30 years) patients are slightly higher than the older (age of >= 30 years) patients (OR=1.2, P = 0.29). The difference is not statistically significant (Kaplan-Meier (KM) curve, Figure 9). Sample size was too low for the greater than five years survival to be considered.

**Table 6.** KM Case Processing Summary data for TNBC patients based on Age and Five Years Survival. (N of Events=survived, Censored=died)

**Case Processing Summary**

| Five_Years_Survival | Age30 | Total N | N of Events | Censored N | Percent |
|---|---|---|---|---|---|
| <= 5 Yr | < 30 | 231 | 189 | 42 | 18.2% |
|  | >= 30 | 20489 | 17289 | 3200 | 15.6% |
|  | Overall | 20720 | 17478 | 3242 | 15.6% |
| > 5 Yr | < 30 | 25 | 17 | 8 | 32.0% |
|  | >= 30 | 2491 | 2128 | 363 | 14.6% |
|  | Overall | 2516 | 2145 | 371 | 14.7% |
| Overall | Overall | 23236 | 19623 | 3613 | 15.5% |

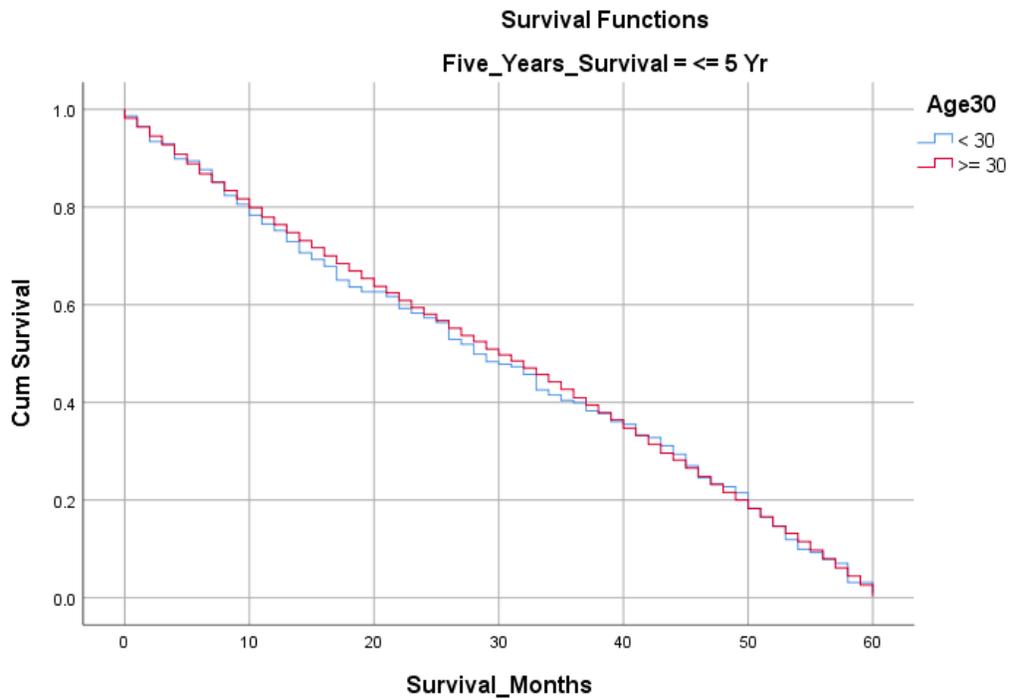

**Figure 9**. KM survival months for TNBC patients by Age and Five_Years_Survival <=5 years.

# Younger TNBC patients' survival by NPI Prognosis

Studies on breast cancer generally use the term prognosis for patients as being excellent, good, and/or poor based on pathological information including tumor size, tumor grade, lymph nodes involvement and other prognostic factors. However, there has been no report on the comparison of the patients' prognosis based on age with respect to the prognostic group (i.e., excellent prognosis, good prognosis, etc.) that the patients belonged to.

In our experiments, we classified TNBC patients based on their NPI prognosis scheme in order to better understand the significance of the younger age on patient's prognosis.

Examining Table 7, for the Moderate NPI prognosis, we find that the odds of survival for the younger (age of < 30 years) patients are slightly higher than the older (age of >= 30 years) patients (OR=1.6, P = 0.036, Figure 10), and the difference is statistically significant. As for the poor NPI prognosis, the younger patients have better survival rates than the older patients; however, the difference is not statistically significant (OR=1.25, P=0.50, Figure 11). Sample sizes for the excellent and good prognoses were too low to be considered.

**Table 7**. KM Case Processing Summary data for TNBC patients based on NPI_Prognosis and Age. (N of Events=survived, Censored=died)

**Case Processing Summary**

| NPI_Prognosis | Age30 | Total N | N of Events | Censored N | Percent |
|---|---|---|---|---|---|
| Excellent | < 30 | 8 | 6 | 2 | 25.0% |
|  | >= 30 | 810 | 636 | 174 | 21.5% |
|  | Overall | 818 | 642 | 176 | 21.5% |
| Good | < 30 | 9 | 6 | 3 | 33.3% |
|  | >= 30 | 1983 | 1635 | 348 | 17.5% |
|  | Overall | 1992 | 1641 | 351 | 17.6% |
| Moderate | < 30 | 130 | 101 | 29 | 22.3% |
|  | >= 30 | 12090 | 10210 | 1880 | 15.6% |
|  | Overall | 12220 | 10311 | 1909 | 15.6% |
| NA | < 30 | 44 | 39 | 5 | 11.4% |
|  | >= 30 | 3868 | 3298 | 570 | 14.7% |
|  | Overall | 3912 | 3337 | 575 | 14.7% |
| Poor | < 30 | 65 | 54 | 11 | 16.9% |
|  | >= 30 | 4229 | 3638 | 591 | 14.0% |
|  | Overall | 4294 | 3692 | 602 | 14.0% |
| Overall | Overall | 23236 | 19623 | 3613 | 15.5% |

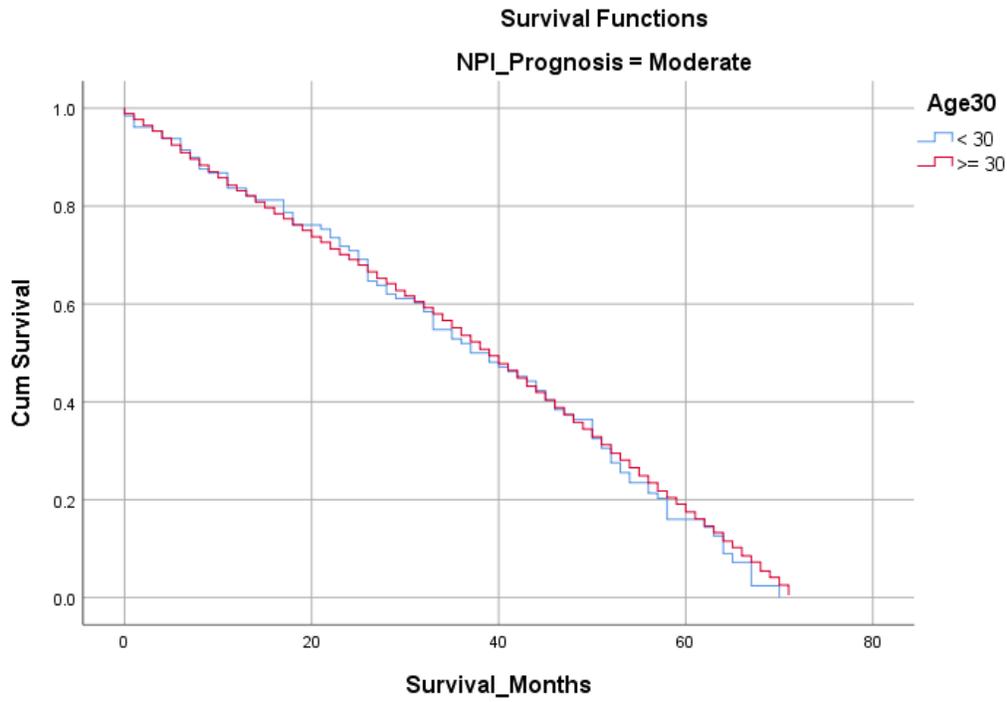

**Figure 10**. KM survival months for TNBC patients by Age and Moderate NPI_Prognosis.

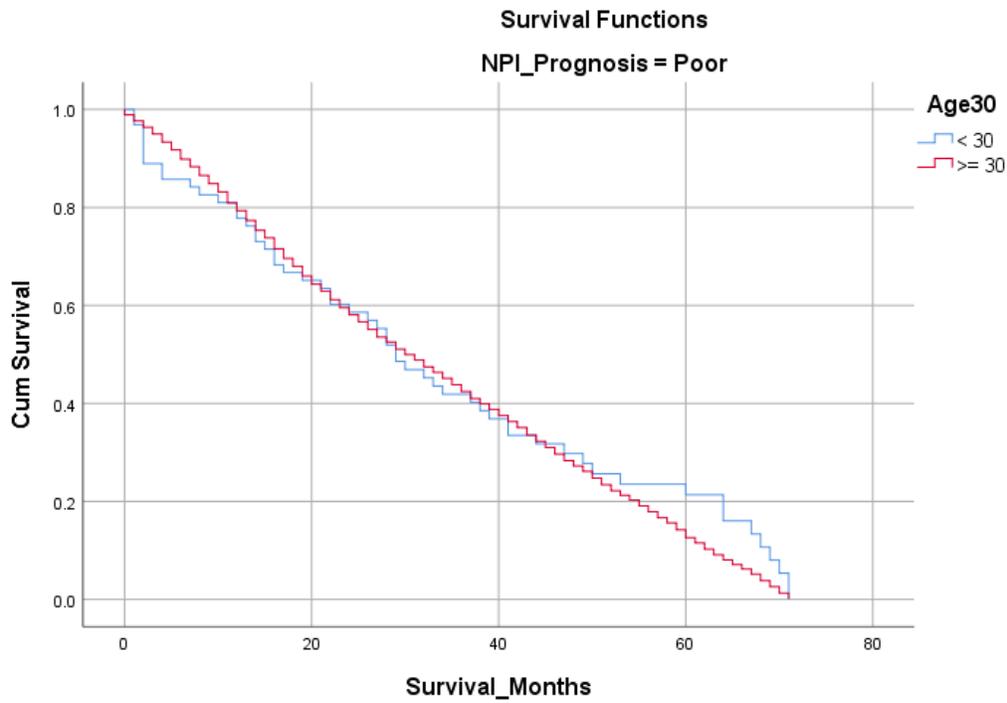

**Figure 11**. KM survival months for TNBC patients by Age and Poor NPI_Prognosis.

## 4.3 Experimental Results Summary

**Non-TNBC patients vs TNBC patients (if this is a subtitle of 4.3 the font size must be 12 )**

The experimental results from this study confirm that non-TNBC patients have higher odds of survival than the TNBC patients. Regarding the age factor, both younger (< 30 years) and older (>= 30 years) non-TNBC patients doing better than the TNBC counterparts.

**Analysis of TNBC Patients**

The experimental results suggest that the TNBC patients have different odds of survival with respect to their age, ethnicity/race groups, Five_Years_Survival, and overall NPI prognosis. Experiments have been analyzed based on ethnicity_race groups HWhite, NHAPI, NHWhite, and NHBlack. Other ethnicity_race groups have not been considered due to low samples. In the following, we highlight the differences among TNBC patients:

1) The younger age alone had no significant impact on the overall survival rate of the TNBC patients.

2) Considering age and ethnicity_race, the older and younger HWhite, NHWhite, NHAPI, and NHBlack had no statistically significant differences in the survival rates.

3) Marital status had no significance on the survival of younger TNBC.

4) For two years and five years survival, the differences between younger and older TNBC patients were not statistically significant.

5) Age had no significant impact on the survival rate for the NPI prognosis results. Both younger and older patients had similar results in all categories of NPI Prognosis.

In summary, we could conclude that younger age had no significance on the prognosis and survival rate of the TNBC patients. It should be noteworthy to mention that the low sample sizes may have contributed to the outcome that has been found, i.e., being inconclusive, regarding the significance of younger age on the overall survivability of the TNBC patients. Such assertion can only be verified with higher sample sizes for the TNBC experimental analyses.

## 5. Conclusion and Future Direction

Breast cancer prognostication is a vital element for providing effective treatment for breast cancer patients. Existing studies suggest that TNBC patients tend to have worse prognosis compared to non-TNBC counterparts. Given the poor prognosis of TNBC, cancer-related outcomes must be estimated accurately. Many factors responsible for the poor clinical outcomes observed in TNBC, including age, race/ethnicity, grade, tumor size, lymph node status, among others. Available research data are not conclusive enough to make a convincing argument for or against a biological or clinical difference in TNBC patients based on these factors.

This study was designed to study the effects of the younger age on breast cancer survivability among TNBC patients utilizing population-based Surveillance, Epidemiology, and End Results (SEER) data to confirm whether younger age had any prognostic significance.

Our experimental results on SEER dataset confirm the existing research reports that TNBC patients have worse prognosis compared to non-TNBC based on age. Our main goal was to investigate whether younger age has any significance on the survivability of TNBC patients. Experimental results do not show that younger age has any significance on the prognosis and survival rate of the TNBC patients.

# Future direction

The experiments in this study showed little differences in the survivability of the younger TNBC patients compared to the older patients. However, the differences were not statistically significant. Experiments on a bigger TNBC data set may give us a more conclusive answer.

Current medical literature suggests that factors such as food habits, work/occupational environment, genetics and family history, obesity, and access to health care could influence the survivability rate. Additional experiments should be conducted based on these factors to determine their impact on the prognosis in connection with younger age.